\documentclass[twocolumn,aps,prl,showpacs]{revtex4}
\usepackage{epsfig}
\usepackage{graphics}

\usepackage{epsfig}
\newcommand{\hc}{\hat{c}}

\begin{document}

\title{Optimized Double-well quantum interferometry with Gaussian-squeezed states}
\author{Y. P.  Huang}
\author{M. G. Moore}
\affiliation{Department of Physics \& Astronomy, Michigan State
University, East Lansing, MI 48824}


\begin{abstract}
A Mach-Zender interferometer with a gaussian number-difference
squeezed input state can exhibit sub-shot-noise phase resolution
over a large phase-interval. We obtain the optimal level of
squeezing for a given phase-interval $\Delta\theta_0$ and particle
number $N$, with the resulting phase-estimation uncertainty smoothly
approaching $3.5/N$ as $\Delta\theta_0$ approaches $10/N$, achieved
with highly squeezed states near the Fock regime. We then analyze an
adaptive measurement scheme which allows any phase on
$(-\pi/2,\pi/2)$ to be measured with a precision of $3.5/N$
requiring only a few measurements, even for very large $N$. We
obtain an asymptotic scaling law of $\Delta\theta\approx
(2.1+3.2\ln(\ln( N_{tot}\tan\Delta\theta_0)))/N_{tot}$,  resulting
in a final precision of $\approx 10/N_{tot}$. This scheme can be
readily implemented in a double-well Bose-Einstein condensate
system, as the optimal input states can be obtained by adiabatic
manipulation of the double-well ground state.
\end{abstract}

\maketitle

Measuring an arbitrary phase with precision well above the standard
quantum limit (SQL), i.e., below shot-noise, has been a
long-standing challenge in quantum interferometry \cite{YurMcCKla86,
HolBur93, BouKas97,GioLloMac04}. The SQL minimum phase uncertainty
is $1/\sqrt{N_{tot}}$, but the theoretical lower limit to the phase
uncertainty, known as the Heisenberg limit (HL), is $1/N_{tot}$,
where $N_{tot}$ is the total number of particles used in the
determination of the phase.
There have been many proposals to achieve $1/N_{tot}$ scaling in a two-input
interferometer, which are based on number-difference squeezed input
states \cite{YurMcCKla86,HolBur93, BouKas97,EckBruPou06}, coherent and/or squeezed vacuum input states \cite{Cav81,DenBscFre06,PezSme08}, or the
maximally-entangled $N$-particle NOON state,
$\frac{1}{2}(|N,0\rangle+|0,N\rangle)$
\cite{BolItaWin96,HueMacPel97,Ger00,MunNemMil02}. Recently, the double-well Bose-Einstein
condensate (BEC) has emerged as a promising system for high-precision matter-wave
 interferometry \cite{SchHofAnd05,GatAlbFol06,Lee06,PezSmerBer06,JoShiWil07}, including progress towards
atom-counting at the single-particle level \cite{ChuSchMey05,
SchHopPer05}. For this system, the squeezed-vacuum protocols are not
applicable, while the NOON state is not suited to determine an
unknown phase due to the periodicity of the phase-distribution
\cite{PezSme05,HraReh05}. This leaves number-difference squeezed
states as a viable candidate, although to date there has been no
systematic study of how to measure arbitrary phases at or near HL
precision in the large-$N$ limit. In this Letter we perform such an
analysis and show that an asymptotic scaling of $(\ln(\ln
N_{tot}))/N_{tot}$ can be achieved via multiple adaptive
measurements with Gaussian number-squeezed states, which can be
readily created in a double-well BEC.

For measuring a phase of $\theta=0$, it has been shown that the
Twin-Fock (TF) state, and the related Pezze-Smerzi (PS) state can
achieve Heisenberg scaling
\cite{KimPfiHol98,HraReh05,PezSme06,UysMey07}. We find, however, for
$\theta\neq 0$  the phase-uncertainty of the TF and PS states
rapidly decay to worse-than-SQL, and in the limit of large $N$
approach constant values, independent of $N$. While the PS state was
only investigated for $\theta=0$, Kim {\it et al} \cite{KimPfiHol98}
investigated $\theta\neq 0$ for the TF state with $N=100$. They
claim a phase uncertainty $\sim 1/N_{tot}$ for $\theta<1/N$, and
growing rapidly thereafter. Our results similarly indicate that the
TF and PS states become worse than shot-noise for $\theta\gg 1/N$.

To find the optimal input state, we
constrain ourselves to the ground-states of a double-well BEC with repulsive interactions for experimental obtainability.
The double-well BEC system is described by the Hamiltonian,
\begin{equation}
\label{Bose-Hubbard}
    \hat{H}(\chi)=-2\tau \hat{J}_x+\delta\hat{J}_z+U \hat{J}^2_z,
\end{equation}
where $\tau$ is the inter-well tunneling rate, $U$ is the atom-atom
interaction strength, and $\delta$ is the asymmetric tilt of the
double-well, presumably due to the external perturbation being
measured. The angular momentum operators are defined as
$\hat{J}_x=\frac{1}{2}(\hat{c}^\dagger_L\hat{c}_R+\hat{c}^\dagger_R\hat{c}_L)$,
$\hat{J}_y=\frac{1}{2i}(\hat{c}^\dagger_L\hat{c}_R-\hat{c}^\dagger_R\hat{c}_L)$
and
$\hat{J}_z=\frac{1}{2}(\hat{c}^\dagger_L\hat{c}_L-\hat{c}^\dagger_R\hat{c}_R)$,
with $\hc_L,\hc_R$ being the annihilation operators for particles in
the two localized modes. For repulsive atom-atom interaction, and
$\delta\approx 0$ the ground state is very close to a Gaussian
squeezed (GS) state of the form
$|\sigma\rangle\propto\sum_{n=-N/2}^{N/2}e^{-n^2/4\sigma^2}|n\rangle$,
where $|n\rangle$ is a number-difference eigenstate satisfying
$\hat{J}_z|n\rangle=\ n|n\rangle$. The width $\sigma$, depends on
the parameter $u=U/\tau$, and is given by $\sigma^2=N/4\sqrt{1+u N}$
\cite{ImaLewYou97}. The nature of our adaptive measurement scheme
requires that we tune $u$ to an optimal value which takes into
account our prior knowledge of $N$ and $\theta$, thus we have $u\to
u(N,\theta)$. This tuning is accomplished by varying $U/\tau$ via a
Feshbach resonance and/or changing the shape of the double-well
potential, and allows $\sigma$ to be varied between $0$ and
$\sqrt{N}/2$, corresponding to maximal number-difference squeezing
and no squeezing, respectively.

To implement a Mach-Zehnder interferometer (MZI), we set $U=0$ and
allow tunneling for $t=\pi/4\tau$ duration, thus realizing a
linear 50/50 beamsplitter, described by the propagator
$e^{i\frac{\pi}{2}\hat{J}_x}$. This is followed by a sudden raising
of the potential barrier to turn off tunneling and allow phase
acquisition due to the small but non-vanishing $\delta$. Holding the
system for a measurement time $T$, a phase shift of $\theta=-\delta
T$ will be acquired, described by the propagator
$e^{i\theta\hat{J}_z}$. The barrier is then lowered again to implement a second beamsplitter.

In the MZI, a symmetric input state $|\Psi_{in}\rangle$ is
transformed into the phase-dependent output state
$|\Psi_{out}\rangle=e^{i\frac{\pi}{2} \hat{J}_x}e^{i\theta\hat{J}_z}
e^{i\frac{\pi}{2} \hat{J}_x}|\Psi_{in}\rangle= e^{-i\theta
\hat{J}_y}e^{i\pi\hat{J}_x}|\Psi_{in}\rangle$ \cite{YurMcCKla86}.
Applying this transformation to a typical GS state, results in an
output state whose properties are easily understood using the Bloch
sphere quasi-probability distribution \cite{EckBruPou06}, which
assigns a probability to each point on a sphere of radius $N/2$
according to $P(\theta,\phi)=|\langle
N/2|e^{i\hat{J}_y(\pi/2-\theta)}e^{i\hat{J}_z\phi}|\Psi_{out}\rangle|^2$.
Note any $N$-atoms state  for which all atoms are in the same
single-particle state can be written in the form
$e^{-i\hat{J}_z\phi}e^{-i\hat{J}_y(\pi/2-\theta)}|N/2\rangle$, which
has eigenvalue $N/2$ with respect to the projection of $\vec{J}$
onto the axis defined by $\theta,\phi$. In this picture, the TF and
PS states are thin equatorial rings, and a GS input state is an
ellipse centered on the $J_x$ axis, compressed along the
$J_z$-direction. Typical GS input and output states are shown in
figure 1(a).

Phase information is obtained by measuring the number difference
between the two interferometer modes, which projects the output
state onto a $\hat{J}_z$ eigenstate. Quantum fluctuations in this
measurement are governed by the projection of the output
distribution onto the $J_z$-axis. Due to the rigid rotation, the
width of the projection will be determined by a $\theta$-dependent
combination of the $J_z$ and $J_x$ noise of the input distribution.
The goal of this paper is thus to find the optimal amount of
squeezing to minimize the phase uncertainty given a fixed particle
number $N$ and an initial estimated phase $\theta_0$ with
uncertainty $\Delta\theta_0$.

\begin{figure}
\epsfig{figure=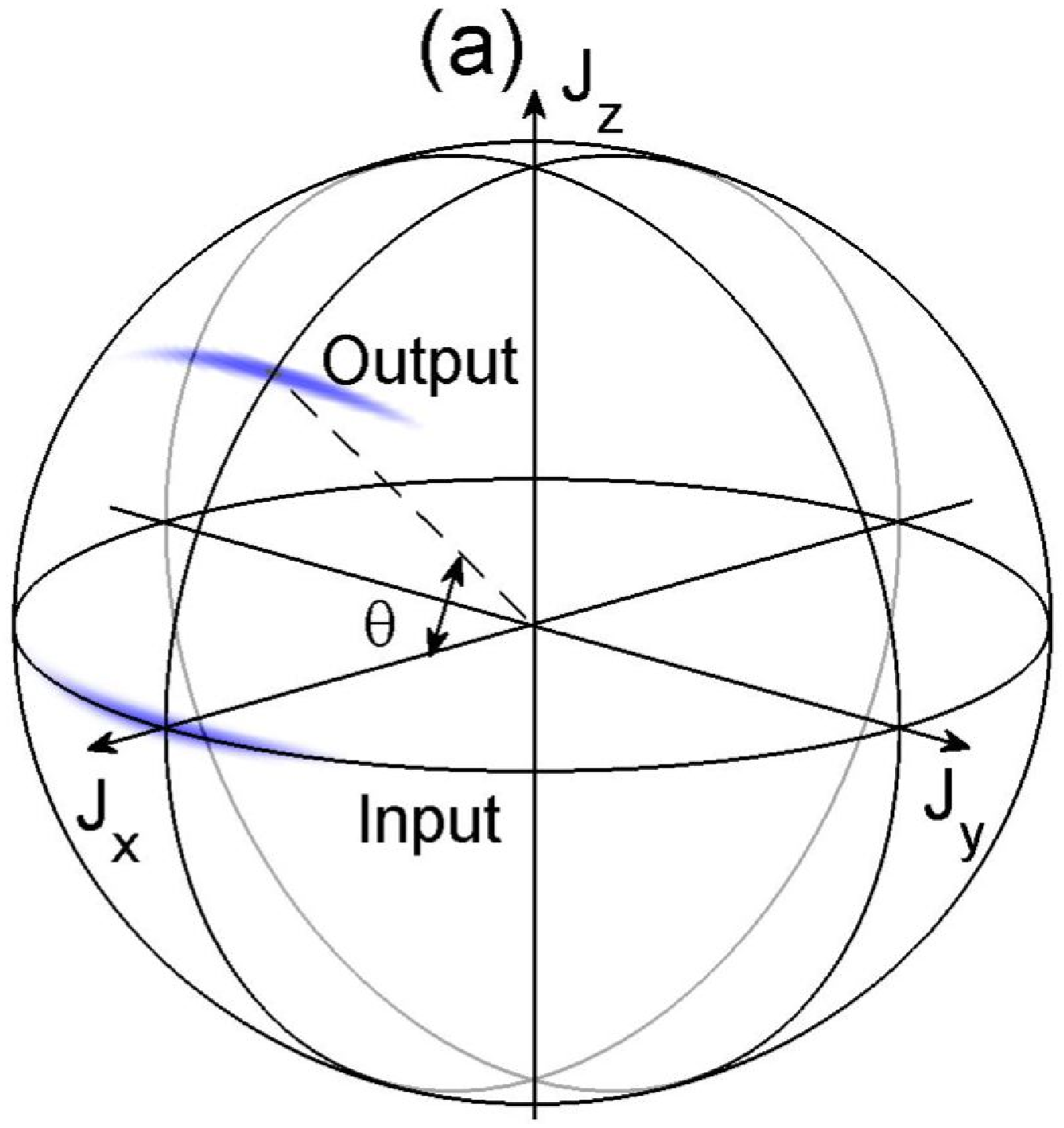, width=4.2cm}
\epsfig{figure=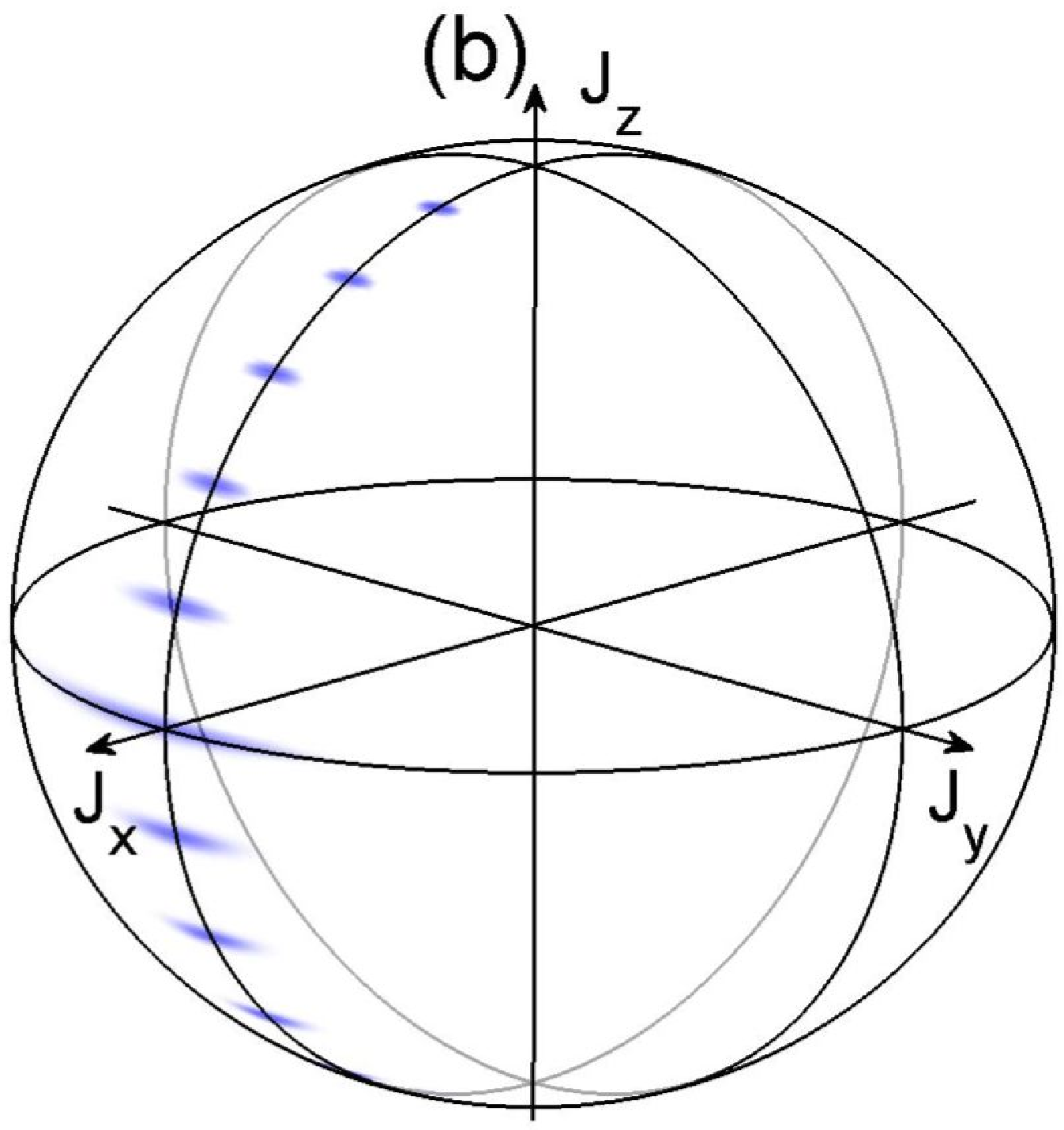, width=4.2cm}
\epsfig{figure=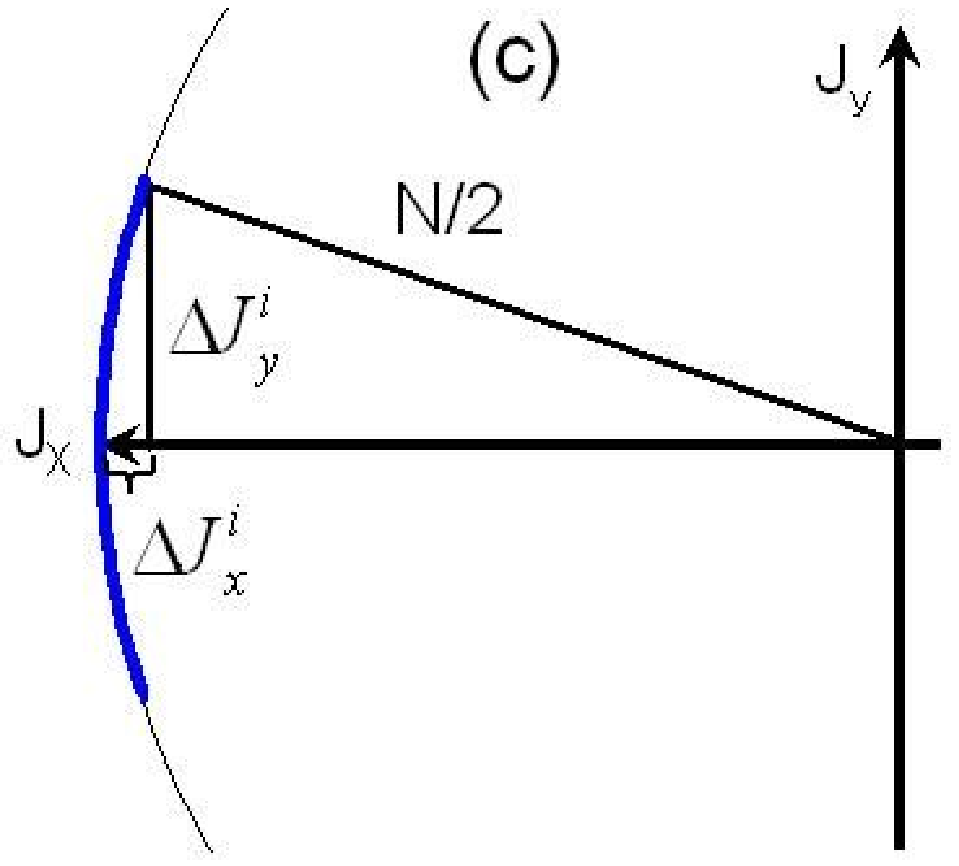, width=3.8cm}
\epsfig{figure=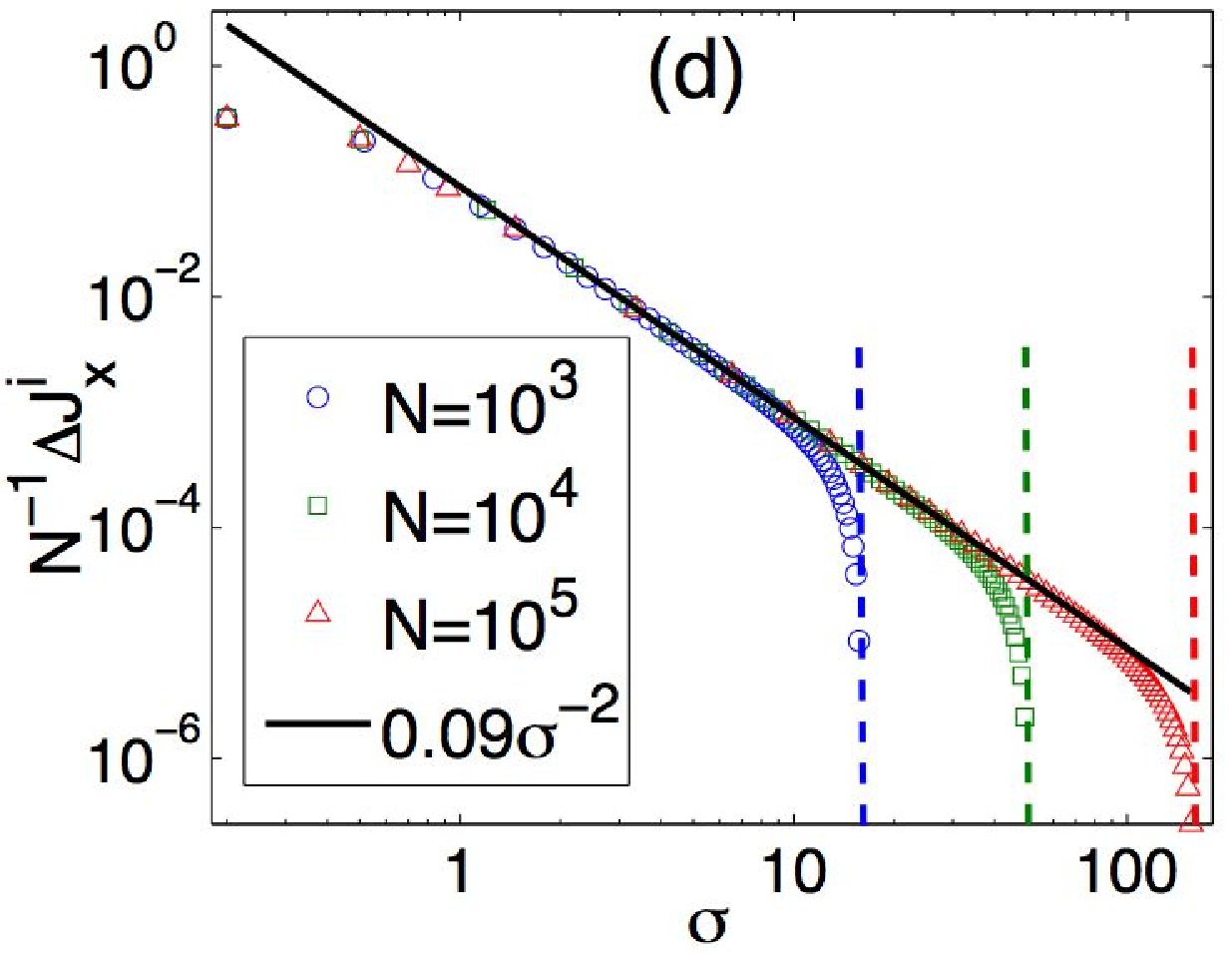, width=4.6cm}
\caption{(Color online) Bloch-sphere analysis of the MZI with a GS input state: (a) A typical GS state at input and output stages; (b) output states for optimized input states with $\theta=0,\pm\frac{\pi}{12},\pm\frac{\pi}{6},\pm\frac{\pi}{4},\pm\frac{\pi}{3},\pm\frac{5\pi}{12}$; (c) geometric origin of the $J_x$ input noise, $\Delta J_x^i$; (d) numerical results plotting $\Delta J_x^i/N$ versus $\sigma$ on a log-log scale for three different $N$-values, validating the functional form of $\Delta J_x^i$ derived geometrically from (c). The dashed vertical lines correspond to $\sigma=\frac{\sqrt{N}}{2}$, where $\Delta J_x^i$ drops to zero.
\label{fig1}}
\end{figure}

Before we present numerical results from a rigorous Bayesian
analysis, we first use linearized error propagation to provide an
approximate analytical description of the interferometer
performance. An analytical result is important for predicting the
behavior at large $N$, where a numerical result is
inaccessible.  In this approach, the phase uncertainty
is estimated by evaluating
$\Delta\theta=[\partial\langle\hat{J}_z\rangle/\partial\theta]^{-1}
\Delta J_z$ at the interferometer output. For input
states symmetric around $n=0$, the expectation
values at the output are related to those at the input via $\langle \hat{J}_z
\rangle=\sin\theta \langle \hat{J}^i_x \rangle$ and $\Delta
J_z=\sqrt{\cos^2\theta \Delta {J^i_z}^2+\sin^2\theta
 \Delta {J^i_x}^2}$.  For GS
states, $\langle \hat{J}^i_x \rangle\approx N/2$, and $ \Delta
J^i_z=\sigma$, which immediately leads to $\langle
\hat{J}_z\rangle=N(\sin\theta)/2$. In figure 1(c) we see that $\Delta J^i_x\approx N/2-\sqrt{N^2/4- \Delta
{J^i_y}^2}$. Since GS state is a
minimum uncertainty state with $\Delta J^i_y \Delta J^i_z =\langle
\hat{J}^i_x \rangle/2$, we see that $ \Delta J^i_y =N/4\sigma$. This leads to
$\Delta J^i_x=\alpha N/\sigma^2$, with $\alpha\approx 0.06$. Exact numerical calculations verify this analytic form for $1 \ll \sigma \ll \sqrt{N}/2$, as shown in figure 1(d), but with
$\alpha=0.09$. Inserting these results into the error-propagation
formula, we find
\begin{equation}
\label{Deltatheta}
    \Delta\theta\approx\frac{2\sigma}{N} \sqrt{1+\left[\frac{0.09
    N\tan\theta}{\sigma^3}\right]^2}.
\end{equation}
The TF and PS states roughly correspond to a fixed $\sigma\lesssim
1$, resulting in $\lim_{N\to\infty}\Delta\theta =
.18\tan\theta/\sigma^2$, which quickly becomes saturated to an
$N$-independent constant for $\theta\gg 1/N$, a result we have
verified numerically with exact Bayesian calculations. On the other
hand, if holding $u$ fixed so that $\sigma\sim N^{1/4}$, the phase
uncertainty scales as $\Delta\theta\sim 1/N^{3/4}$ for $\theta=0$,
as discussed in \cite{PezColSme05}, and would eventually saturate to
$\sim 1/\sqrt{N}$ for $\theta\neq 0$. Rather than holding $\sigma$
or $u$ fixed, we propose varying $u$ and thus $\sigma$ with $N$ in
order to minimize the phase variance. By setting
$d\Delta\theta/d\sigma = 0$ we find
\begin{eqnarray}
    \sigma_{min}(\theta,N)&\approx& .503 (N\tan|\theta|)^{1/3},\label{sigmamin}\\
    \Delta\theta_{min}(\theta,N)&\approx& 1.23(\tan|\theta|)^{1/3}/N^{2/3}\label{Deltathetamin}
\end{eqnarray}
From self-consistency, these expressions are
valid only when $10/N\lesssim |\theta| \lesssim
\tan^{-1}(.137 \sqrt{N})\approx \pi/2$.

We now employ rigorous Bayesian analysis to quantify the phase
uncertainty and validate our approximate analytic results, again
assuming that $|\theta|$ is not too close to $\pi/2$. According to
Bayes theorem, upon a measurement result $n_m$, the probability that
the actual phase is $\phi$ is $P(\phi|n_m)=P(n_m|\phi)/\int d\theta
P(n_m|\theta)$, where $P(n|\theta)=|\langle n|e^{-i\theta
\hat{J}_y}|\psi_{in}\rangle|^2$. The error-propagation result
(\ref{Deltathetamin}) is very close to the $68\%$ confidence
interval of $P(\phi|n)$ because the underlying number distribution
of the optimized GS output state is well approximated by the
gaussian distribution $P(n|\theta)\approx[\sqrt{2\pi} \Delta
n]^{-1}e^{-(n-N\sin\theta/2)/2\Delta n^2},$ where $\Delta
n=\sqrt{1+\frac{\tan^2\theta}{2\tan^2\theta_a}}\sigma_{min}(\theta_a,N)$,
with $\theta$ being the unknown phase, and $\theta_a$ being the
assumed phase used for optimization. Provided that
$|\theta|\sim|\theta_a|$, the dependence on $\theta$ is weak, and we
have $\Delta n\approx \sqrt{3/2}\sigma_{min}(\theta_a,N)$. This
shows that the most-probable outcome is $\bar{n}=N\sin\theta/2$,
which is sensitive to the sign of $\theta$. Because $\Delta n$ is
only weakly dependent on $\theta$, the inverted distribution
$P(\phi|n)$ will also be close to Gaussian in the small-angle
regime, with width given by (\ref{Deltathetamin}). To make a
theoretical performance analysis for a fixed $\theta$, we average
over all possible measurement outcomes, defining
$P(\phi|\theta)=\sum_{n} P(\phi|n)P(n|\theta)$
\cite{PezSme06,UysMey07}. This can be interpreted as the probability
for an experimenter to infer $\phi$ given a true phase-shift of
$\theta$. The phase uncertainty $\Delta\theta$ is then defined as
the $68\%$ confidence interval, via
$\int_{\theta-\Delta\theta}^{\theta+\Delta\theta} d\phi\,
P(\phi|\theta)=.68$.

Using this approach, together with the exact double-well
ground-state, we numerically find $u_{min}$, the value of $U/\tau$
which minimizes the phase uncertainty. In figure 2(a), we plot the
corresponding $\sigma_{min}=\sigma(u_{min})$ as a function of $N$
for several $\theta$s. Also shown is a least-squares fit to the
$N>10^3$ data (including many data points not shown explicitly) to
the analytic form (\ref{sigmamin}), giving
\begin{eqnarray}
\label{sigmin}
 \sigma_{min}(\theta,N)=\left\{
               \begin{array}{c}
                 1.00, ~~~~~~~~~~~~~~~~~~~~~~~|\theta|<10/N; \\
                 0.45 (N\tan|\theta|)^{1/3}, ~~~~~|\theta|>10/N,
               \end{array}\right.
\end{eqnarray}
in good agreement with our analytical result. Inverting Eq.
(\ref{sigmin}) to leads to
\begin{eqnarray}
\label{umin}
 u_{min}(\theta,N)=\left\{
               \begin{array}{c}
                 \frac{N}{16}-\frac{1}{N}, ~~~~~~~~~~~~~~~~~~~~|\theta|<10/N; \\
                 \frac{1.52}{(\tan|\theta|)^{4/3}N^{1/3}}-\frac{1}{N}, ~~~~~|\theta|>10/N.
               \end{array}\right.
\end{eqnarray}

\begin{figure}
\epsfig{figure=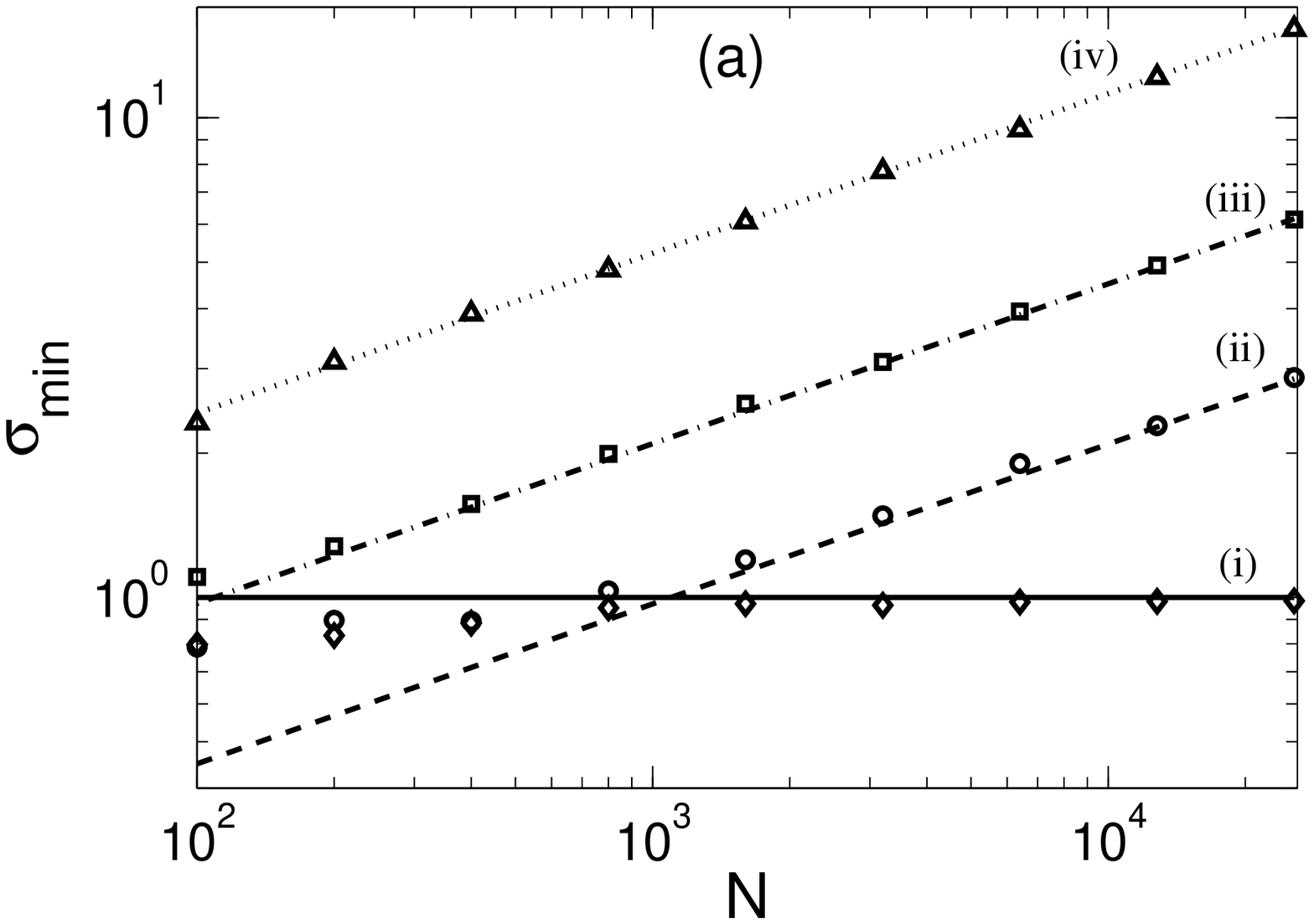, width=8.0cm} \epsfig{figure=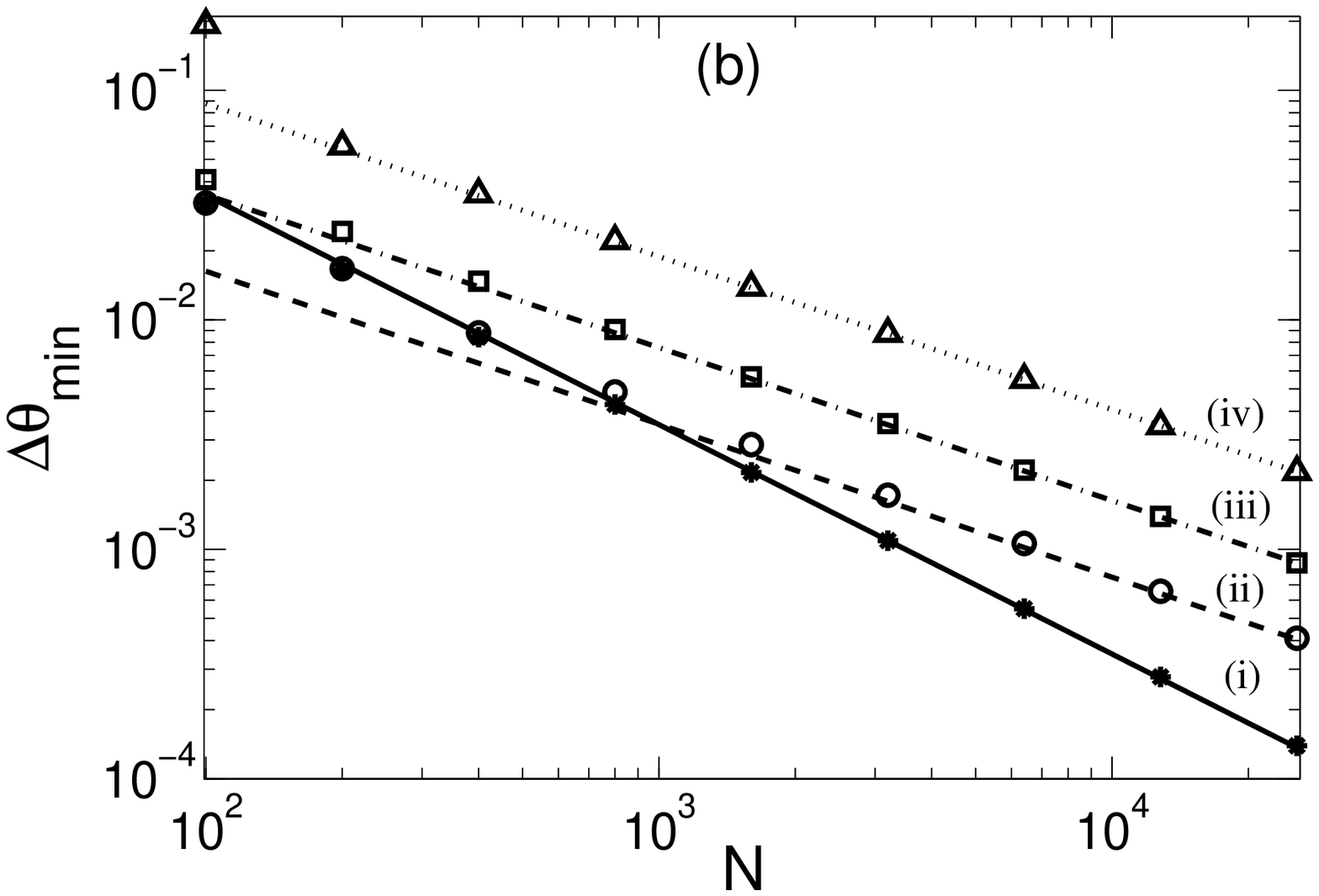,
width=8.0cm} \caption{Figure (a): Optimal width $\sigma_{min}$
versus $N$ for different $\theta$; (b): corresponding minimized
phase uncertainty $\Delta\theta_{min}$. In both figures, from (i) to
(iv) the interferometer phases are $\theta=0,~0.01,~0.1$ and $1$.
The data points represent numerical results from strict Bayesian
analysis using the exact ground states of a double-well BEC, while
the straight lines represent the asymptotical forms of
Eq.(\ref{sigmin}) and Eq.(\ref{dthmin}), respectively. \label{fig2}}
\end{figure}

In Fig. \ref{fig2}(b) we plot the corresponding minimized
$\Delta\theta_{min}$ versus $N$ for several phases, achieved by
setting $\sigma=\sigma_{min}(\theta,N)$.
Again fitting the $N>10^3$ data to the analytic form of (\ref{Deltathetamin}), we find
\begin{eqnarray}
\label{dthmin}
 \Delta\theta_{min}(\theta,N)=\big\{
               \begin{array}{c}
                 3.50/N, ~~~~~~~~~~~~~~~~~~~~|\theta|<10/N; \\
                 1.63 (\tan|\theta|)^{1/3}/N^{2/3}, ~~|\theta|>10/N. \\
               \end{array}
\end{eqnarray}
The difference between the prefactor here and (\ref{Deltathetamin}) is primarily due to a factor of approximately $\sqrt{2}$ which comes from the definition of $P(\phi|\theta)$.

In practice, $\theta$ is not known a-priori, hence it is not clear
what value for $\theta$ to use in determining
$\sigma_{min}(\theta,N)$ via Eq. (\ref{sigmin}). If we assume prior
knowledge of the form $P(\theta)\propto
\exp[-(\theta-\theta_0)^2/2\Delta\theta_0^2]$, we should first
remove $\theta_0$ by adding $\theta_0/T$ to the tilt $\delta$ during
phase acquisition, and then use $\sigma_{min}(\Delta\theta_0,N)$.
After obtaining a measurement result $n_1$, the estimated
uncertainty $\Delta\theta_1$ is then be determined via
$\int_{\theta_1-\Delta\theta_1}^{\theta_1+\Delta\theta_1}d\theta'
P(\theta'|n_1)$, with $P(\theta'|n_1)$ being given by Bayes theorem.
This will result in
$\Delta\theta_1\sim\Delta\theta_{min}(\Delta\theta_0,N)$. Based on
Eq. (\ref{dthmin}), this uncertainty appears to scale only as
$N^{-2/3}$, only a slight improvement of $N^{1/6}$ over the SQL.
However, in many applications requiring high precision, the phases
are very small, in which case the phase uncertainty can be reduced
considerably  due to the explicit phase dependence in
(\ref{dthmin}). This is in contrast to a shot-noise-limited
interferometer,  where $\Delta\theta=1/\sqrt{N}$ for all $\theta$s
not too close to $\pm \pi/2$. The explicit theta-dependence in the
optimized scheme
 is due to the fact that stronger number-squeezing
can be tolerated at smaller angles before the $\hat{J}^i_x$ noise
becomes detrimental. For example, if the phase is known to be
smaller than $1/\sqrt{N}$, we have
$\Delta\theta_{min}\le1.63/N^{5/6}$, which is now an $N^{1/3}$
improvement over the SQL. As can be seen from Eq. (\ref{dthmin}), a
maximum sensitivity of $3.5/N$ can be achieved for $|\theta|<10/N$,
which is true Heisenberg scaling.

In fact, almost any phase between $-\pi/2$ and $\pi/2$ can be
measured at the maximum precision of  $3.5/N$ if the present scheme
is combined with multiple adaptive measurements
\cite{YurMcCKla86,DenBscFre06}. After the first measurement as
described above, we can again rebalance the interferometer by adding
$\theta_1/T$ to the tilt, followed by a second measurement with
$\sigma_2=\sigma_{min}(\Delta\theta_1,N)$, with result $n_2$. The
Bayesian distribution for $\theta_2$ will then be approximately
$P(\theta_2|n_2,n_1)\propto
\exp[-(\theta_2-\bar{\theta}_2)^2/2y_2^2]$, where
$\bar{\theta}_j=\sin^{-1}(2n_j/N)$ and $1/y_j^2=\sum_{k=0}^j
1/\Delta\theta_k^2$. Since $\Delta\theta_j$ is much smaller than
$\Delta\theta_{j-1}$ we can say $y_j\approx \Delta\theta_j\sim
\Delta\theta_{min}(\Delta\theta_{j-1},N)$. In other words, since the
distribution after a measurement is much narrower than the previous
distribution, multiplying the distributions has little effect, so
that the final uncertainty is effectively determined by the
resolution of the final measurement alone. After $M$ iterations,
with $\sigma_j=\sigma_{min}(\Delta\theta_{j-1},N)\approx 0.57 \left(
N\tan\Delta\theta_0/2.1\right)^{3^{-j}}$, we find
\begin{equation}
    \Delta\theta_M \sim (2.1/N)\left( N\tan\Delta\theta_0/2.1\right)^{3^{-M}}.\label{DtM}
\end{equation}
While the above expressions are good estimates of the expected
behavior, in practice each $\sigma_j$ and $\Delta\theta_j$ would be
computed exactly by applying Bayes theorem after each measurement.
This procedure should be repeated only until $\Delta\theta_M\lesssim
10/N$, after which an addition measurement will push the phase
uncertainty to $3.5/N$. The final measurement is then made using the
GS state with $\sigma=1$, which lies at the edge of the
maximally-squeezed Fock regime defined by $\sigma \ll 1$.  Thus an
arbitrary phase can be measured at $3.5/N$ precision with $M+1$
measurements in total. Setting $\Delta\theta_M=10/N$, and solving
for $M$ gives
\begin{equation}
\label{M}
    M\approx0.9\ln(\ln(N\tan\Delta\theta_0/2.1))-0.4,
\end{equation}
where again this is just an estimate subject to run-to-run fluctuations.
For $\theta_r=\pi/3$ and $N=10^4$,
$M=1.6$, i.e. only 2 or 3 total measurements will be required. For
 $N=10^{12}$, $M=2.6$, requiring 3 or 4 measurements.
Even for $\Delta\theta_0$ extremely close to $\pi/2$, $M$ remains
small, for example, $\Delta\theta_0=\pi/2-10^{-10}$ gives $M\approx
2.7$ for $N=10^4$, and $M\approx 3.1$ for $N=10^{12}$. Hence, for
arbitrary phases in $(-\pi/2,\pi/2)$, our interferometer converges
quickly to the $3.5/N$ precision within a few measurements,
regardless of $N$. The final experimental value for the initial
unknown phase is then $\theta=\sum_{j=1}^{M+1}\theta_j$ with a
quantum-limited uncertainty of $\Delta\theta=3.5/N$. The total
number of atoms used to obtain this precision is $N_{tot}=(M+1)N$.
For large enough $N$, we can approximate
$\ln(N_{tot}/2(M+1))\approx\ln(N_{tot})$, so that $
    M+1\approx 0.9 \ln(\ln(N_{tot}\tan\Delta\theta_0))+0.6,
$
which leads to the asymptotic scaling law
\begin{equation}
\label{limit}
    \Delta\theta\approx (2.1+3.2\ln(\ln( N_{tot}\tan\Delta\theta_0)))/N_{tot}.
\end{equation}
That the scaling law should depend on the initial phase interval has been previously pointed out \cite{DurDow07}.
As our final approximation effectively overestimates $N_{tot}$,  the uncertainty approaches (\ref{limit}) from {\it below} as $N$ increases.

\begin{figure}
\epsfig{figure=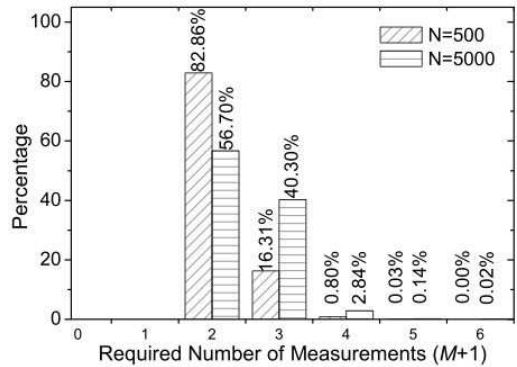, width=7.0cm} \caption{Monte-Carlo
simulation results showing the percentage of runs which achieved the
maximum precision of $3.5/N$ after $(M+1)$ measurements, plotted
versus $(M+1)$. \label{fig4}}
\end{figure}

In order to verify the accuracy of Eq. (\ref{M}), as well as
determine the magnitude of the run-to-run fluctuations, we have
carried out exact Monte-Carlo simulations of many measurements of
the phase $\pi/6$, with an initial uncertainty
$\Delta\theta_0=\pi/3$. During each simulation run, the measurement
outcome was randomly selected according to the output distribution,
and the phase information was determined numerically via Bayes
theorem. The prescribed measure-rebalance process was iterated until
the estimated phase uncertainty reaches $3.5/N$. Figure \ref{fig4}
shows the percentage of runs which attain the desired resolution on
the $(M+1)$th iteration, for two different $N$ values, with $10^4$
runs each.  The averages are $\overline{M+1}=2.2$ for $N=500$, and
$2.5$ for $5000$. Equation (\ref{M}) gives $2.2$ and $2.5$ as well.
The corresponding variances $\Delta(M+1)$ are $0.4$ and $0.6$. We
note that as the average approaches a half-integer value, the
minimum possible variance approaches $0.5$, because $M$ is
constrained to integer values. Thus the fluctuations are close to
minimum allowed values.

For TF and PS states, adaptive measurement schemes are extremely inefficient. This is due in part to their inability to distinguish positive from negative phases, which makes rebalancing impossible.  But even if this were overcome, the primary difficulty is that the phase-uncertainty is $N$-independent for large phases, so that $\sim N^2$ measurements are required to obtain $1/N$ precision. This results in $1/N_{tot}^{1/3}$ scaling, worse than SQL.

The above discussions have assumed that the input state is optimally
squeezed to width $\sigma_{min}$. A realistic input state, however,
may deviate from $\sigma_{min}$, due to imprecise control over $u$
and/or imprecise knowledge of $N$. A straightforward error analysis
shows that our scheme is extremely robust against such
uncertainties. The increase in the single-measurement phase
uncertainty $\delta\Delta\theta$ due to fluctuations in $u$ and $N$
is found to be
\begin{equation}
    \delta\Delta\theta=
    \left|\frac{\partial \Delta\theta(N,u)}{\partial N} \right|\delta N+
    \left|\frac{\partial^2 \Delta\theta(N,u_)}{\partial^2 u}
    \right|\frac{1}{2}\delta u^2
\end{equation}
evaluated at $u=u_{min}$. The scaling with $\delta u^2$ reflects the
fact that $u=u_{min}$ is a local minimum with respect to the phase
uncertainty. This gives
\begin{equation}
    \delta\Delta\theta/\Delta\theta_{min}=(2/3)\delta N/N+(1/8)(\delta u/u_{min})^2,
\end{equation}
so that a $10\%$ variation in $N$ leads to a $7\%$ variation in the
phase uncertainty, while even a $100\%$ uncertainty in $u$ only
results in a $13\%$ variation. For our purposes, these increases are
essentially negligible, and are independent of the values of $N$ or
$\theta$. Of course there are many other potential sources of error,
e.g. the precision with which the tilt can be rebalanced, and the
precision with which the scattering length can be set to zero during
interferometer operation. Reaching the Heisenberg limit in a
double-well BEC interferometer will clearly require major
technological advances in many areas. Assuming that a level of
precision significantly below the SQL is eventually obtained, the
scheme we have developed will be the optimal method to obtain this
precision, whether or not it is close to the Heisenberg limit.

In conclusion, we have shown that an adaptive GS state scheme has
three advantages over previously discussed MZ interferometry
schemes. It (1) can readily be implemented in a double-well BEC
system, (2) can achieve a resolution well beyond the SQL for a wide
range of phases with a single measurement, and (3) quickly converges
to a final precision $\approx10/N_{tot}$ with only a few
adaptive measurements.

This work is supported in part by Nation Science Foundation Grant
No. PHY0653373.

\end{document}